\documentclass[manuscript]{aastex631}

\usepackage{hyperref}

\shorttitle{GLEs in Cycle 24}
\shortauthors{Liu et al.}

\begin{document}

\title{Effects of Coronal Magnetic Field Configuration on Particle Acceleration and Release during the Ground Level Enhancement Events in Solar Cycle 24}

\correspondingauthor{Xiangliang Kong}
\email{kongx@sdu.edu.cn}

\author[0000-0002-9515-9406]{Wenlong Liu}
\affiliation{School of Space Science and Physics, Institute of Space Sciences, Shandong University, Weihai, Shandong 264209, People's Republic of China}
\affiliation{Yunnan Key Laboratory of Solar Physics and Space Science, Kunming 650216, People's Republic of China}

\author[0000-0003-1034-5857]{Xiangliang Kong}
\affiliation{School of Space Science and Physics, Institute of Space Sciences, Shandong University, Weihai, Shandong 264209, People's Republic of China}
\affiliation{Institute of Frontier and Interdisciplinary Science, Shandong University, Qingdao, Shandong 266237, People's Republic of China}

\author[0000-0003-4315-3755]{Fan Guo}
\affiliation{Los Alamos National Laboratory, Los Alamos, NM 87545, USA}
\affiliation{New Mexico Consortium, Los Alamos, NM 87544, USA}

\author[0000-0003-3936-5288]{Lulu Zhao}
\affiliation{Department of Climate and Space Sciences and Engineering, University of Michigan, Ann Arbor, MI 48109, USA}

\author[0000-0002-9634-5139]{Shiwei Feng}
\affiliation{School of Space Science and Physics, Institute of Space Sciences, Shandong University, Weihai, Shandong 264209, People's Republic of China}

\author[0000-0002-1576-4033]{Feiyu Yu}
\affiliation{School of Space Science and Physics, Institute of Space Sciences, Shandong University, Weihai, Shandong 264209, People's Republic of China}

\author[0000-0002-4520-2170]{Zelong Jiang}
\affiliation{School of Space Science and Physics, Institute of Space Sciences, Shandong University, Weihai, Shandong 264209, People's Republic of China}

\author[0000-0001-6449-8838]{Yao Chen}
\affiliation{School of Space Science and Physics, Institute of Space Sciences, Shandong University, Weihai, Shandong 264209, People's Republic of China}
\affiliation{Institute of Frontier and Interdisciplinary Science, Shandong University, Qingdao, Shandong 266237, People's Republic of China}

\author[0000-0002-0850-4233]{Joe Giacalone}
\affiliation{Department of Planetary Sciences, University of Arizona, Tucson, AZ 85721, USA}

\begin{abstract}

Ground level enhancements (GLEs) are extreme solar energetic particle (SEP) events that are of particular importance in space weather.
In solar cycle 24, two GLEs were recorded on 2012 May 17 (GLE 71) and 2017 September 10 (GLE 72), respectively, by a range of advanced modern instruments.
Here we conduct a comparative analysis of the two events by focusing on the effects of large-scale magnetic field configuration near active regions on particle acceleration and release.
Although the active regions both located near the western limb, temporal variations of SEP intensities and energy spectra measured in-situ display different behaviors at early stages.
By combining a potential field model, we find the CME in GLE 71 originated below the streamer belt, while in GLE 72 near the edge of the streamer belt.
We reconstruct the CME shock fronts with an ellipsoid model based on nearly simultaneous coronagraph images from multi-viewpoints, and further derive the 3D shock geometry at the GLE onset.
The highest-energy particles are primarily accelerated in the shock-streamer interaction regions, i.e., likely at the nose of the shock in GLE 71 and the eastern flank in GLE 72, due to quasi-perpendicular shock geometry and confinement of closed fields.
Subsequently, they are released to the field lines connecting to near-Earth spacecraft when the shocks move through the streamer cusp region.
This suggests that magnetic structures in the corona, especially shock-streamer interactions, may have played an important role in the acceleration and release of the highest-energy particles in the two events.
\end{abstract}

\keywords{Solar energetic particles (1491), Solar particle emission (1517), Solar coronal mass ejections (310), Shocks (2086), Solar coronal streamers(1486)}

\section{Introduction} \label{sec:intro}
During solar flares and coronal mass ejections (CMEs), a large number of charged particles are accelerated to high energies (up to $\sim$10 MeV for electrons and $\sim$GeV for ions) near the Sun, known as solar energetic particles \citep[SEPs; see reviews by][]{1999SSRv...90..413R,2016LRSP...13....3D}. When SEPs propagate through the interplanetary medium and arrive at the vicinity of the Earth, they can have hazardous radiation impacts on the Earth's space environment \citep{2009SSRv..147..187V,2012SSRv..171..161S}.
Understanding the origin, acceleration, and transport of SEPs remains an outstanding question in solar and space weather physics.
Ground level enhancement (GLE) events are a category of extremely intense SEP events and of particular interest to space weather. In GLE events, the highest-energy particles can reach $\sim$GeV and penetrate into Earth’s atmosphere, producing secondary particles that can be detected by ground-based neutron monitors (NMs). Recent observations have indicated that GLE events are closely associated with fast CMEs and high-energy particles are primarily accelerated at CME-driven shocks near the Sun \citep[e.g.][]{2009ApJ...706..844R,2009ApJ...693..812R,2012SSRv..171...23G}.

Although the diffusive shock acceleration mechanism has been widely accepted for GLE and large SEP events, it remains unclear where the highest-energy particles are produced.
When a CME-driven shock moves through the complex magnetic structures in the corona, both the shock geometry and shock strength can vary significantly along the shock front \citep[e.g.][]{2011ASTRA...7..387P,2015ApJ...799..167K,2016ApJ...833...45R,2019ApJ...876...80K,2021ApJ...913...99K,2022SpWea..2002894J}.
One of the most prominent structures in the corona are streamers, characterized by closed magnetic field and high plasma density, therefore lower Alfv\'en speed than the ambient corona \citep{2013ChSBu..58.1599C,2013ApJ...776...55K}. Consequently, when a CME-shock interacts with nearby streamers, the shock strength can be enhanced owing to a higher Alfv\'enic Mach number and the shock obliquity is dominated by quasi-perpendicular geometry \citep[e.g.,][]{2016ApJ...833...45R,2021ApJ...913...99K,2022SpWea..2002894J}, leading to more efficient particle acceleration \citep{1987ApJ...313..842J,2005ApJ...624..765G}.

Recent numerical simulations have shown that the large-scale coronal magnetic structures can play an important role in the acceleration of SEPs in the early stages \citep[e.g.,][]{2009A&A...507L..21S,2012ApJ...753...87K,2013ApJ...778...43K,2015ApJ...810...97S,2017ApJ...851...38K,2019ApJ...883...49K,2022ApJ...941...23C,2022FrASS...901429K,2022ApJ...925L..13Y}.
\citet{2017ApJ...851...38K,2019ApJ...883...49K} numerically modeled particle acceleration at coronal shocks propagating through ambient streamers and found that the highest-energy ($>$100 MeV) particles are concentrated around the shock-streamer interaction region due to the trapping of closed field lines and quasi-perpendicular shock geometry. \citet{2022ApJ...925L..13Y} further showed that the forms of particle energy spectra can vary significantly along the shock front, exhibiting either a double power-law or a power-law with exponential rollover.
Observations of multi-spacecraft SEP events have also shown that the higher-energy SEPs are more spatially confined and the energy spectral shapes measured at different locations can differ largely for the same event \citep[e.g.,][]{2013AIPC.1539..116M,2017ApJ...843..132C,2021SoPh..296...36B}.

In observations, type II radio bursts have been used as indicators of energetic electrons accelerated at CME-shocks. Close associations are often found between type II radio bursts and shock-streamer interactions \citep[e.g.,][]{2003ApJ...590..533R,2004A&A...413..363M,2011A&A...530A..16C,2012ApJ...750..158K,2013SoPh..282..543S,2015ApJ...798...81K,2013ApJ...767...29F,2017SoPh..292...74S,2019ApJ...871..212F,2019A&A...624L...2M,2020ApJ...893..115C,2021ApJ...913...99K,2023SoPh..298...59S}.
Recently, it has been shown that streamers or streamer-like coronal structures can also affect the acceleration, trapping and release of energetic protons or heavy ions in large SEP events or GLEs \citep[e.g.,][]{2016ApJ...833...45R,2017ApJ...839...79K,2020ApJ...903...41C,2022ApJ...926..227F}.
\citet{2017ApJ...839...79K} analyzed two GLE events in solar cycle 23 and suggested that the accelerated relativistic protons may be trapped within large-scale magnetic loops and released when the CME reaches the tips of coronal streamers or due to CME-induced magnetic reconnection.
\citet{2020ApJ...903...41C} speculated that shock-streamer interaction possibly can provide the favorable condition for the occurrence of the 774 AD SEP event, the largest in the last 10,000 year.
For the SEP event on 2013 June 21, \citet{2022ApJ...926..227F} showed that the acceleration of high-energy SEPs to above 100 MeV by the subcritical CME-shock can be explained by the prolonged shock-streamer interaction, which is favorable for high acceleration efficiency as simulated by \citet{2017ApJ...851...38K,2019ApJ...883...49K}.

During solar cycle 24, only two GLE events are recorded, i.e., the events on 2012 May 17 (GLE 71) and 2017 September 10 (GLE 72).
The two events have been intensively studied from different aspects in previous works, including the CMEs, CME-driven shocks and type II radio bursts \citep[e.g.,][]{2013ApJ...763..114S,2016ApJ...833...45R,2018ApJ...853L..18Y,2019ApJ...871....8L,2019NatAs...3..452M}, X-ray and microwave emissions \citep{2020NatAs...4.1140C}, the Fermi gamma-ray emission \citep[e.g.,][]{2020ApJ...893...76K,2021ApJ...915...12K}, and the SEP events \citep[e.g.,][]{2013ApJ...765L..30G,2013ApJ...770...34L,2018SpWea..16.1616C,2016ApJ...818..169D,2018A&A...612A.116B,2018ApJ...863L..39G,2018RAA....18...74Z,2019SpWea..17..419B,2021ApJ...919..146L,2021ApJ...921...26Z,2022ApJ...924..106D}.
In this paper, we present a comparative analysis of the two GLE events by focusing on the effects of larges-scale magnetic field configuration near the active regions on the acceleration and release of the first-arriving ($\sim$GeV) particles.
In Section 2, we present the observations and results, including the difference in temporal variations of SEP intensity profiles and energy spectra, imaging observations of the solar eruptions and 3D reconstruction of the shock geometry, shock-streamer interactions, and magnetic connection between the SEP sources and the Earth observer.
Conclusions and discussion are given in Section 3.

\section{Observations and Results}

\subsection{Temporal Evolution of SEP Intensities and Energy Spectra}
\begin{figure}
\centering
\includegraphics[width=0.95\linewidth]{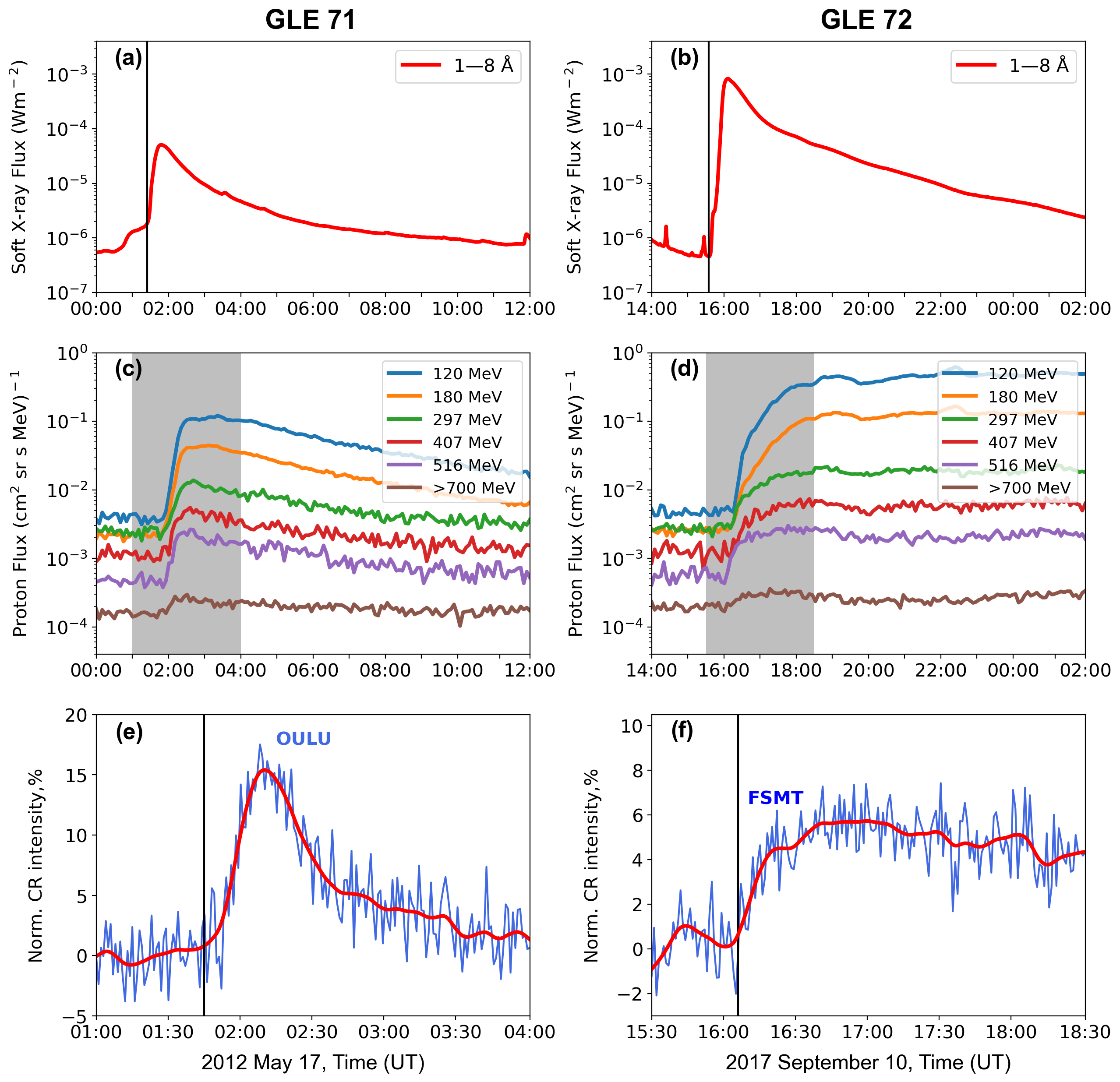}
\caption{(a$-$b) GOES 1$-$8 \AA ~SXR light curves for the two GLE events, GLE 71 on 2012 May 17 and GLE 72 on 2017 September 10. The vertical solid lines indicate the flare start time.
(c$-$d) Proton intensity profiles in different energy channels observed by EPEAD and HEPAD onboard GOES 15.
(e$-$f) Count rates recorded by NM stations from OULU for GLE 71 and FSMT for GLE 72. The red lines display the smoothed data and the vertical solid lines indicate the identified GLE onset time.
}
\label{fig:sepflux}
\end{figure}

Figures \ref{fig:sepflux}(a) and (b) show the GOES 1$-$8 \AA ~soft X-ray (SXR) light curves for the two GLE events, with the starting time marked by vertical solid lines. In the GLE 71 event, the M5.1 flare started at 01:25 UT and peaked at 01:47 UT. In the GLE 72 event, the X8.2 flare started at 15:35 UT and peaked at 16:06 UT.
In both events, the associated flares and CMEs originated near the western limb of the Sun, from NOAA active regions AR11476 (N13W87) and AR12673 (S09W91), respectively. Therefore, the SEP source regions close to the Sun are generally well connected to the Earth, resulting in a significant flux increase of high-energy particles as detected by the instruments at 1 AU both in space and on the ground.

For the analysis of energetic particles, we use the data from two instruments onboard the GOES 15 spacecraft, the Energetic Proton, Electron, and Alpha Detector (EPEAD) and High Energy Proton and Alpha Detector (HEPAD), which measure the protons from 2.5 MeV to \textgreater700 MeV in 11 different energy channels. We consider the calibration of high-energy proton channels as in \citet{2017SpWea..15.1191B} to improve the reliability of differential energy spectra.
Figures \ref{fig:sepflux}(c) and (d) display the temporal evolution of the proton flux in the two events, including the energy channels from P6 to P11 (120 MeV to \textgreater700 MeV).
In both events, the proton intensity exhibits obvious increase in all energy channels.
However, the difference in their time-intensity profiles can also be noticed.
Specifically, taking the 120 MeV channel as an example, in GLE 71, the proton intensity first increased rapidly by more than one order of magnitude within half an hour, then it reached a plateau and kept for $\sim$2 hours, followed by a decay phase.
In comparison, in GLE 72, the proton intensity grows more slowly and takes more than 2 hours before reaching the plateau stage, and the plateau lasts for more than 8 hours.
The proton intensity profile in GLE 71 manifests the typical behavior of well-connected SEP events, while in GLE 72 the Earth observer is not well connected to the source region of high-energy SEPs at the beginning of the event (as shown below in Figure \ref{fig:Bfield2017}).

In GLE events, relativistic particles ($\sim$GeV) can penetrate into the Earth’s atmosphere and produce secondaries that can be detected by NMs on the ground. Due to the difference in the cut-off rigidity, the onset and intensity of recorded GLE events vary with the location of NM stations \citep{2018SoPh..293..136M,2019SoPh..294...22K}.
To identify the GLE onset time, we use the criterion, $f_{onset}$ = $\left \langle f \right \rangle$ + 2$\sigma$, where $\left \langle f \right \rangle$ is the average of background and $\sigma$ is the standard deviation.
For the GLE 71 event, Figure \ref{fig:sepflux}(e) displays the pressure-corrected count rates from the NM stations of Oulu (OULU), where the cut-off rigidity is 0.8 GV.
The data resolution in time is 1 minute and the smoothed data is shown by the red curve.
By using OULU, we then determine the onset time for the GLE 71 event to be 01:44 UT, as indicated by the vertical solid line in Figure \ref{fig:sepflux}(e).
Note that \citet{2013ApJ...765L..30G} also used OULU to determine the onset time and they suggested the onset time was at 01:43 UT when the intensity was at 2$\%$ of the peak.
We assume that the first arriving $\sim$GeV particles propagate along the nominal Parker spiral with a length of 1.2 AU and are scattering-free, therefore they take $\sim$11.3 min from their release site in the corona to the Earth.
To compare with the remote sensing observations in white-light, EUV, and radio wavelengths, the time for electromangetic signals traveling to Earth ($\sim$8.3 min) is subtracted. As a result, the release time of $\sim$GeV particles in the corona is determined to be at 01:41 UT, after normalization to the time of remote sensing signals.

Figure \ref{fig:sepflux}(f) displays the pressure-corrected count rates of Fort Smith (FSMT) for the GLE 72 event, where the cut-off rigidity is 0.3 GV. FSMT shows an early rapid increase compared to other NMs \citep{2019SoPh..294...22K}. The smoothed data of FSMT is plotted as the red curve.
We use FSMT to determine the GLE onset time by applying the same method as mentioned above. The onset time was determined as 16:06 UT, as indicated by the vertical line.
Thus, the release time of GLE particles in the corona was at 16:03 UT, i.e., 3 min earlier after normalization to the time of remote sensing signals.
Note that, by using FSMT as well, the GLE onset time was suggested to be at 16:07 $\pm$ 1 min UT in \citet{2019SoPh..294...22K} and 16:08:30 UT in \citet{2020ApJ...893...76K}. 
In addition, \citet{2018ApJ...863L..39G} took 16:08 UT as the onset time, when the GLE intensity of OULU reached 10\% of its peak.

\begin{figure}
\centering
\includegraphics[width=0.95\linewidth]{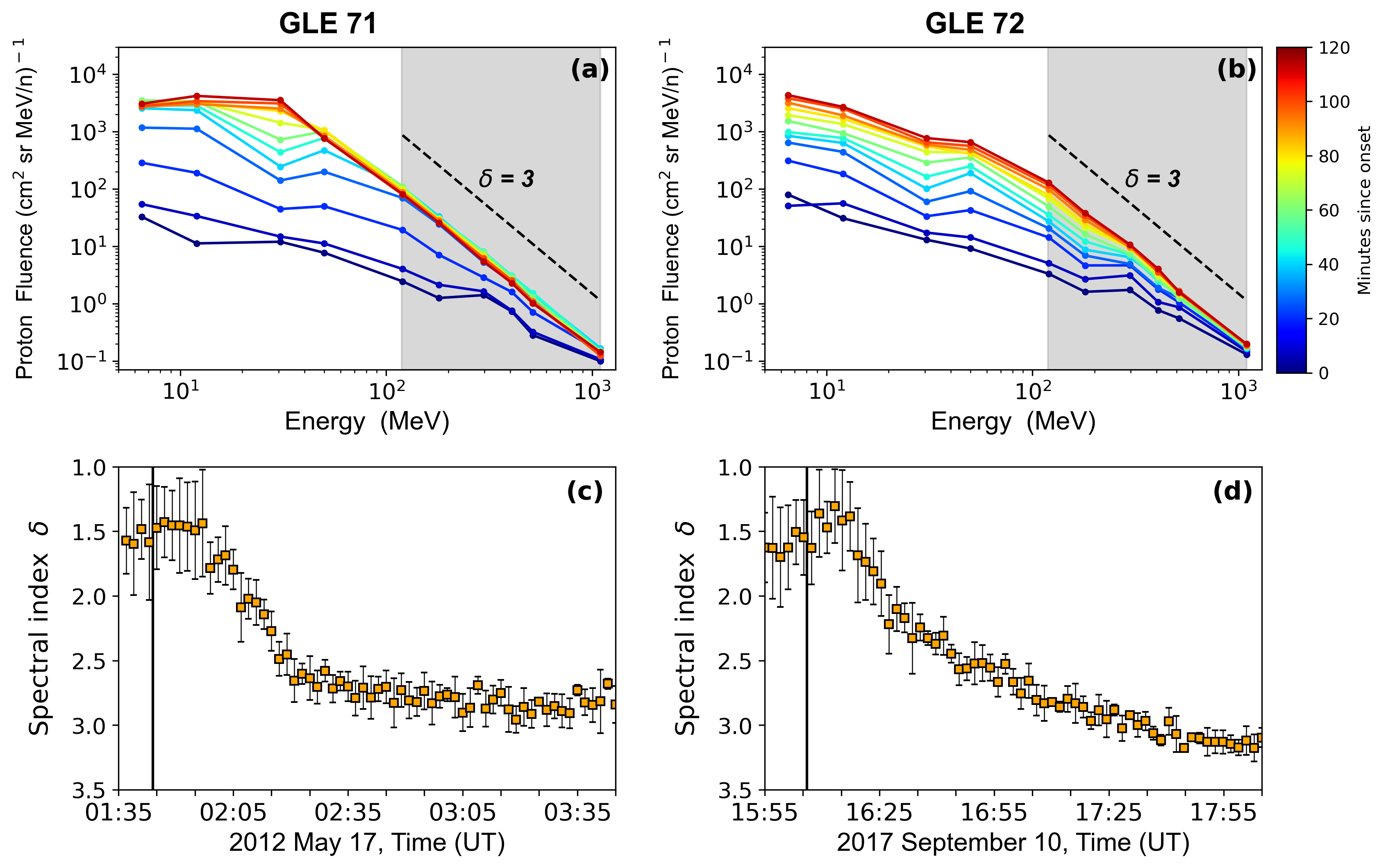}
\caption{(a$-$b) Temporal evolution of the ten-minutes integrated proton energy spectra within the first two hours after the GLE onset time. Data are taken from EPEAD and HEPAD onboard GOES 15.
(c$-$d) Temporal evolution of the spectral indexes at two-minute cadence obtained by fitting the proton energy spectra above 100 MeV with a single power-law function, $F(E) \propto E^{-\delta}$. The vertical solid lines indicate the GLE onset time.
}
\label{fig:sepspec}
\end{figure}

We further analyze the proton energy spectra in the early stages of the two events. Figures \ref{fig:sepspec}(a) and (b) show the temporal evolution of ten-minute integrated energy spectra within the first 2 hours after the GLE onset time, in the energy range between 6.5 MeV and 1009 MeV measured by EPEAD and HEPAD aboard on GOES.
As noted above, we consider the calibration of high-energy proton channels to improve the reliability of differential energy spectra \citep{2017SpWea..15.1191B}.
Overall, in both events, the particle spectra get softer after the GLE onset and the spectral indexes become $\sim$3 after two hours.
However, as a consequence of their difference in the intensity-time profiles, the detailed trends in the variation of spectral slopes at $>$100 MeV appear to be different.
In GLE 71, the energy spectra at high energies ($>$100 MeV) quickly evolve into a power-law distribution within 40 minutes and then remain almost unchanged, while the spectra in GLE 72 steepen slower and continuously in the first 2 hours.
We fit the proton spectra above 100 MeV at a two-minute cadence with a single power-law function, $F(E) \propto E^{-\delta}$, where $E$ is the proton energy and $\delta$ is the spectral index. The fitted spectral indexes with 1-sigma error are plotted as a function of time in the lower panels.
In GLE 71, the spectral index is $\sim$1.45 within the first 14 minutes after the event onset at 01:44 UT, then it increases rapidly to $\sim$2.7 in 20 minutes and remains nearly a constant thereafter.
By contrast, in GLE 72, the spectral index first decreases (the spectrum getting harder) within the first 10 minutes after the onset at 16:06 UT, and then it increases gradually and reaches $\sim$3.2 at 18:00 UT.

The different behaviors as shown in the temporal evolution of SEP intensity profiles and energy spectra indicate that the efficiency of particle acceleration and magnetic connectivity to the SEP sources are likely different in the early stages of the two events.

\begin{deluxetable*}{lcc}
\tablenum{1}
\tablecaption{Summary of timelines (in UT) for the two GLE events\label{tab:event}}
\tablewidth{0pt}
\tablehead{
\colhead{}  & \colhead{GLE 71 (20120517)}  & \colhead{GLE 72 (20170910)}
}
\startdata
Flare start, peak (Class)   &  01:25, 01:47 (M5.1)   &  15:35, 16:06 (X8.2)  \\
Type II burst onset$^a$/Shock formation &  01:32    &  15:53   \\
GeV particle release$^b$ (Shock height)  &  01:41 (2.4 R$_{\sun}$)    &  16:03 (3.7 R$_{\sun}$) \\
GeV particle onset (NM)   &  01:44 (OULU)   &  16:06 (FSMT) \\
CME first appearance in C2  &  01:48  (3.6 R$_{\sun}$)   &  16:00  (3 R$_{\sun}$) \\
\enddata
\tablenotetext{a}{Taken from \citet{2013ApJ...765L..30G} and \citet{2018ApJ...863L..39G}.}
\tablenotetext{b}{Normalized to the time of remote sensing signals at Earth.}
\end{deluxetable*}

\subsection{EUV and White-light Imaging Observations of Solar Eruptions}
To explore the dynamics of CME-driven shocks in the corona, we analyze the EUV and white-light imaging data and reconstruct the 3D geometrical shape of the shock  by fitting multiple-viewpoint observations.

\begin{figure}
\centering
\includegraphics[width=0.85\linewidth]{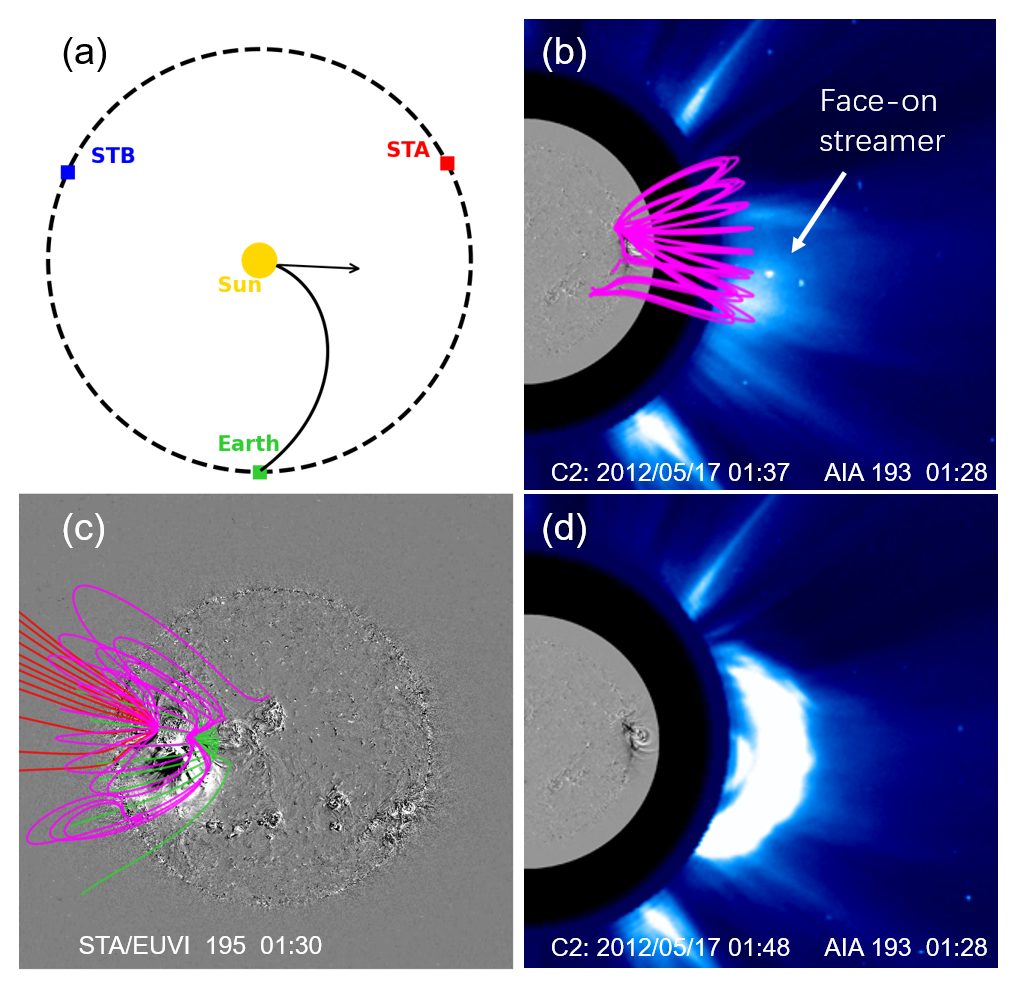}
\caption{(a) Schematics of the locations of Earth (green), STEREO-A (STA, red), and STEREO-B (STB, blue) in the ecliptic plane as seen from the north of the Sun for the GLE 71 event on 2012 May 17.
The arrow indicates the radial direction at the solar flare.
The black curve plots the nominal Parker spiral magnetic field line connecting to the Earth.
(b-d) EUV and white-light images of the CME eruption observed by SDO/AIA, STA/EUVI, and SOHO/LASCO C2 at different times.
The CME feature at early time is shown by EUV running-difference images of SDO/AIA in 193 \AA ~at 01:28 UT, and STA/EUVI in 195 \AA ~at 01:30 UT.
Magnetic field lines obtained from the PFSS model are overplotted in panels (b) and (c). Closed field lines in magenta correspond to the face-on streamer in the FOV of LASCO C2 as indicated by the white arrow in panel (b), and open field lines are plotted in red (positive) and green (negative). The CME first appeared in the FOV of LASCO C2 at 01:48 UT.
}
\label{fig:eruption2012}
\end{figure}

For the GLE 71 event on 2012 May 17, Figure \ref{fig:eruption2012}(a) shows the locations of different observers in the ecliptic plane. The twin STEREO satellites, STEREO-A (STA) and STEREO-B (STB), are both separated from the Earth by approximately 120$^{\circ}$.
The arrow indicates the radial direction at the flare location. The active region AR11476 is located on the western limb  (N13W87) in the view of the Earth, on the eastern hemisphere in the view of STA, and on the backside of the Sun in the view of STB.
As shown in Figures~\ref{fig:eruption2012}(b) and (d), the CME began to move outward at 01:28 UT as observed by SDO/AIA in 193 \AA.
At 01:37 UT, the CME had not yet appeared in the field of view (FOV) of SOHO/LASCO C2. According to the LASCO CME catalog \footnote{\url{https://cdaw.gsfc.nasa.gov/CME_list/}}, the CME moved to 3.61 R$_{\odot}$ at 01:48 UT (Figure~\ref{fig:eruption2012}(d), first appearance in the FOV of LASCO C2), and later it reached 5.43 R$_{\odot}$ at 02:00 UT.
Thus, we can estimate the speed of the outermost CME front as $\sim$1760 km s$^{-1}$.
The magnetic configuration around the active region can be obtained from the potential-field-source-surface (PFSS) model based on the SDO/HMI measurements (see Figure \ref{fig:Bfield2012} below).
As shown in Figures~\ref{fig:eruption2012}(b) and (c), the large-scale closed magnetic field lines represent the streamer (belt) above the active region. It is a face-on streamer as seen in the FOV of LASCO C2.
This indicates that the CME originated below the streamer in the GLE 71 event \citep{2016ApJ...833...45R}. The enhanced brightness at the nose of the CME, as shown in Figure~\ref{fig:eruption2012}(d), may imply the pileup and deflection of the adjacent streamer material as the CME expands and propagates outward.

\begin{figure}
\centering
\includegraphics[width=0.85\linewidth]{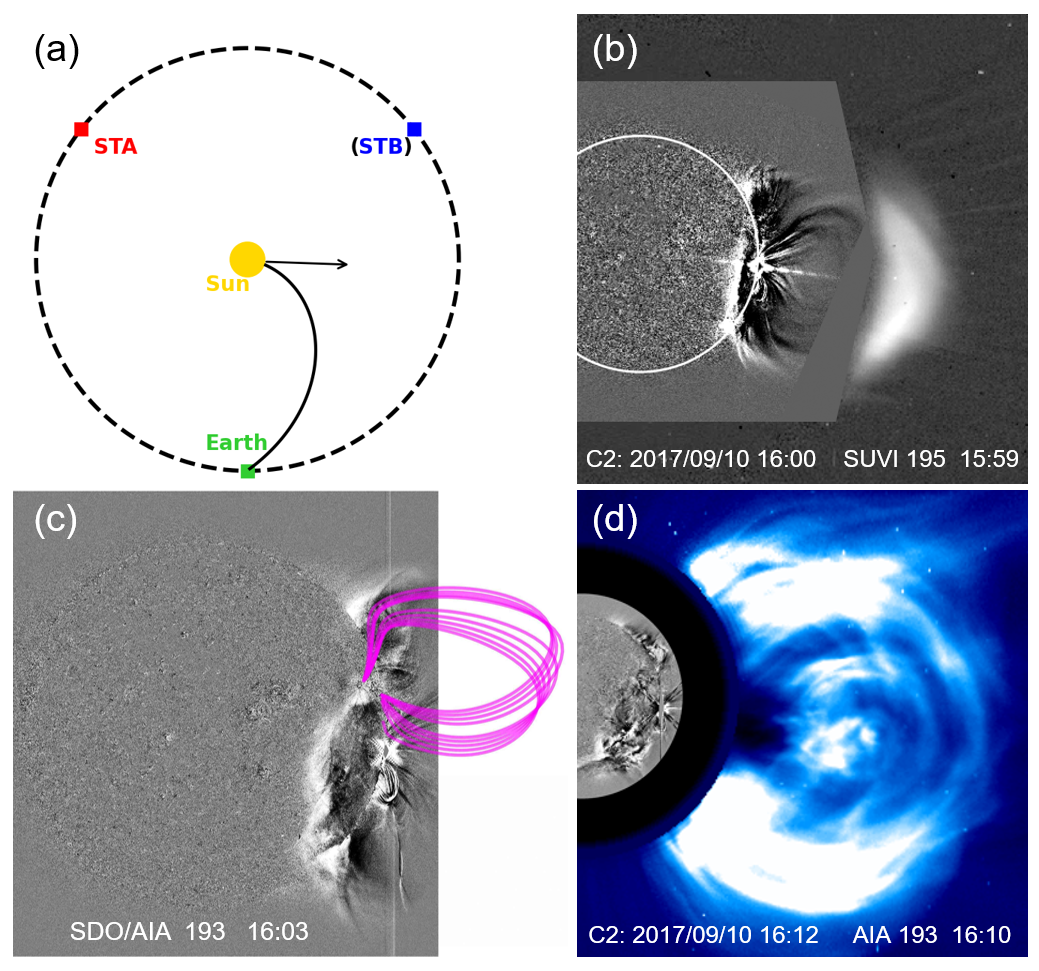}
\caption{(a) Schematics of the locations of Earth (green), STA (red), and STB (blue) in the ecliptic plane as seen from the north of the Sun for the GLE 71 event on 2017 September 10. Communications with STB were lost.
(b) Composite running-difference image observed by GOES-16/SUVI in 195 \AA ~and LASCO C2 at 16:00 UT, when the CME first appeared in the FOV of LASCO C2.
(c) EUV running-difference image observed by AIA 193 \AA ~at GLE particle release time 16:03 UT, with closed field lines in magenta indicating streamer belt located to the northeast of the AR12673.
(d) Composite image of GOES-16/SUVI in 195 \AA ~and LASCO C2 shows the CME at 16:12 UT.
}
\label{fig:eruption2017}
\end{figure}

For the GLE 72 event on 2017 September 10, as shown in Figure \ref{fig:eruption2017}(a), STA and STB are both separated from the Earth by $\sim$128$^{\circ}$. Note that communications with STB were lost. The arrow indicates the radial direction at the flare location, i.e., S09W91 in AR12673 as seen from the Earth. It is a back-side event in the FOV of STA.
As shown in Figures~\ref{fig:eruption2017}(b) and (d), the CME first appeared in the FOV of LASCO C2 at a height of $\sim$3 R$_{\odot}$ at 16:00 UT, and moved to $\sim$6 R$_{\odot}$ at 16:12 UT, indicating the CME speed being $\sim$2900 km s$^{-1}$.
According to the PFSS model, as shown in Figure \ref{fig:Bfield2017} below, the streamer belt is located to the northeast of the AR12673 and above another active region AR12679 \citep{2018SpWea..16..557L}.
Figure~\ref{fig:eruption2012}(c) shows the EUV running-difference image observed by SDO/AIA in 193 \AA ~at the GLE release time at 16:03 UT. The propagation of EUV wave on the solar surface is much slower to the northeast due to strong magnetic field in AR12679 and the streamer belt. Observations of the associated EUV wave in detail can be found in \citet{2018ApJ...864L..24L}.
As shown in Figure~\ref{fig:eruption2012}(d), the presence of bright features at the flanks of the CME may suggest that strong CME-streamer interaction occurred at the CME flanks. This is different from the GLE 71 event.

\begin{figure}
\centering
\includegraphics[width=0.95\linewidth]{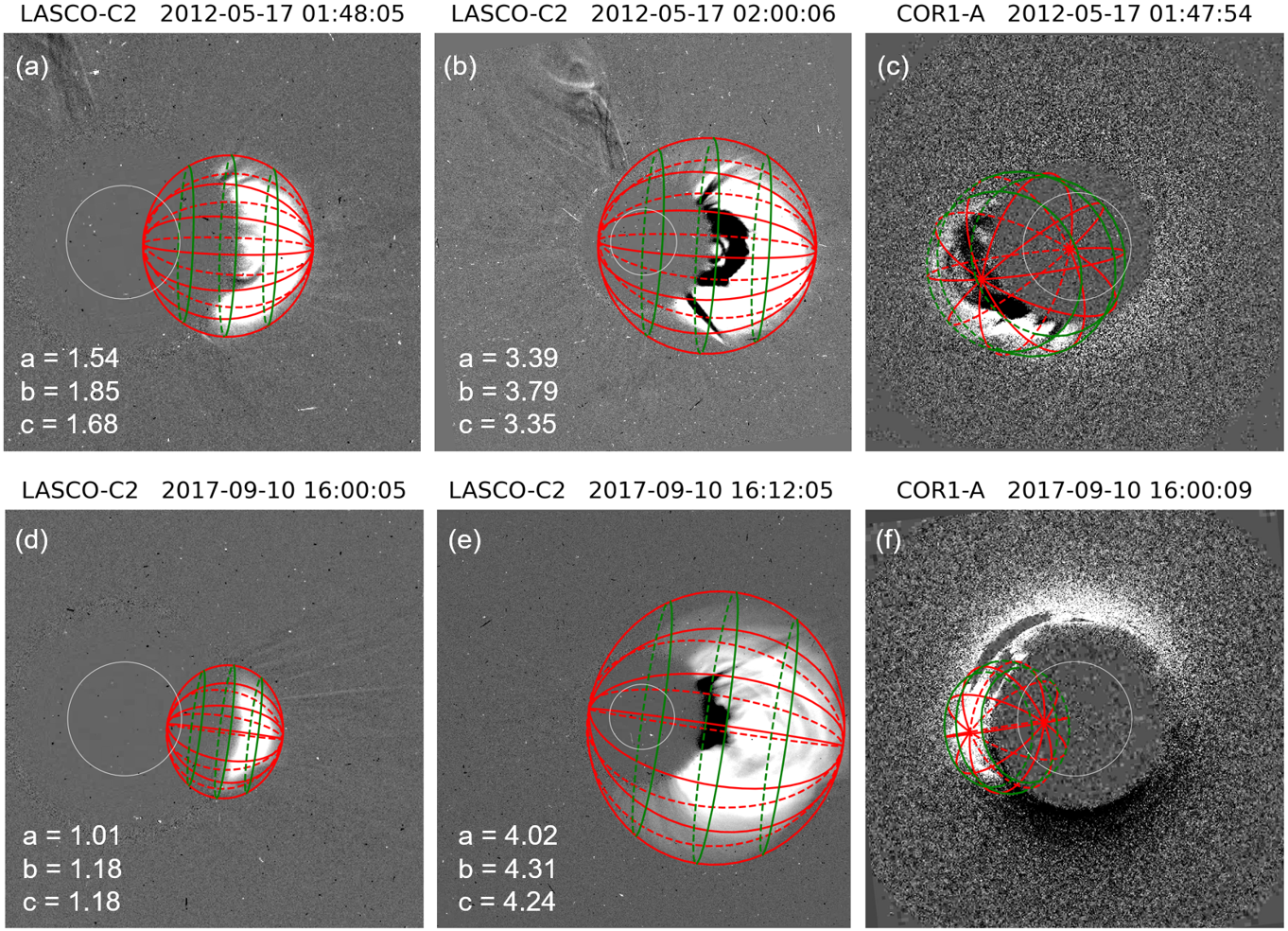}
\caption{Reconstructed 3D shock fronts using the ellipsoid model are overplotted on the running-difference images of LASCO C2 and STA/COR1 for the GLE 71 (a$-$c) and GLE 72 (d$-$f) events.
}
\label{fig:shockfit}
\end{figure}

We reconstruct the 3D shock structure by fitting the coronagraph images from multiple viewpoints using an ellipsoid model \citep{2014ApJ...794..148K}.
The ellipsoid model contains seven geometric parameters, including three parameters for the center of the ellipsoid, three parameters for the lengths of the three principal semi-axes of the ellipsoid, and one for the rotation angle of the ellipsoid.
We obtain the fitting parameters by referring to \texttt{PyThea}, an open-source Python package \citep{2022FrASS...9.4137K}.
Figure \ref{fig:shockfit} displays the fitted shock wave front overlying LASCO C2 and STA/COR1 coronagraph images for the two events.
To better illustrate the shock feature, here we use running-difference images.
Figures~\ref{fig:shockfit}(a) and (d) show the fitting at the time of their first appearance in LASCO C2, and Figures~\ref{fig:shockfit}(c) and (f) show the corresponding modeling from STA/COR1 around the same time.
In the initial stage of the CME-shock, the expansion speed is much larger than the propagation speed of the shock center \citep{2019ApJ...871....8L}.
Therefore, when we perform the fitting for the second images taken by LASCO C2, as shown in Figures~\ref{fig:shockfit}(b) and (e), we only adjust the lengths of the three semi-axes and keep the other four parameters unchanged. The lengths of the three semi-axes, $a$, $b$, and $c$, in units of R$_{\odot}$, are indicated in Figure \ref{fig:shockfit}.
Thus we can calculate the expansion speeds along the three directions.
For example, from the variation of parameter $a$, we calculate the shock speed at the shock leading edge, being 1790 km s$^{-1}$ for GLE 71 and 2910 km s$^{-1}$ for GLE 72. Then, we can further derive the location of the 3D shock front at the GLE particle release time by linear interpolation/extrapolation, as shown in Figures \ref{fig:Bfield2012}(d) and \ref{fig:Bfield2017}(d) below.

\subsection{Magnetic Field Configuration in the Corona and Magnetic Connectivity to the Earth}

To obtain the large-scale magnetic field configuration around the active regions, we carry out the PFSS extrapolation using the SDO/HMI magnetogram data as input.
Here we use \texttt{pfsspy}, the open-source package for Python.
The height of source surface is set to be 2.5 R$_{\odot}$ in both events.
Since magnetic connection between the SEP source region and the observer is critical to understand the onset time and intensity profiles of SEP measurements at 1 AU, we trace the interplanetary magnetic field (IMF) lines from the Earth back to the source surface at 2.5 R$_{\sun}$.
We assume that the IMF lines follow the nominal Parker spirals, based on the average solar wind speed measured 6 hours prior to the eruption at 1 AU.
By utilizing the PFSS solution, we can further trace the magnetic field lines from the source surface down to the photosphere.

\begin{figure}
\centering
\includegraphics[width=0.95\linewidth]{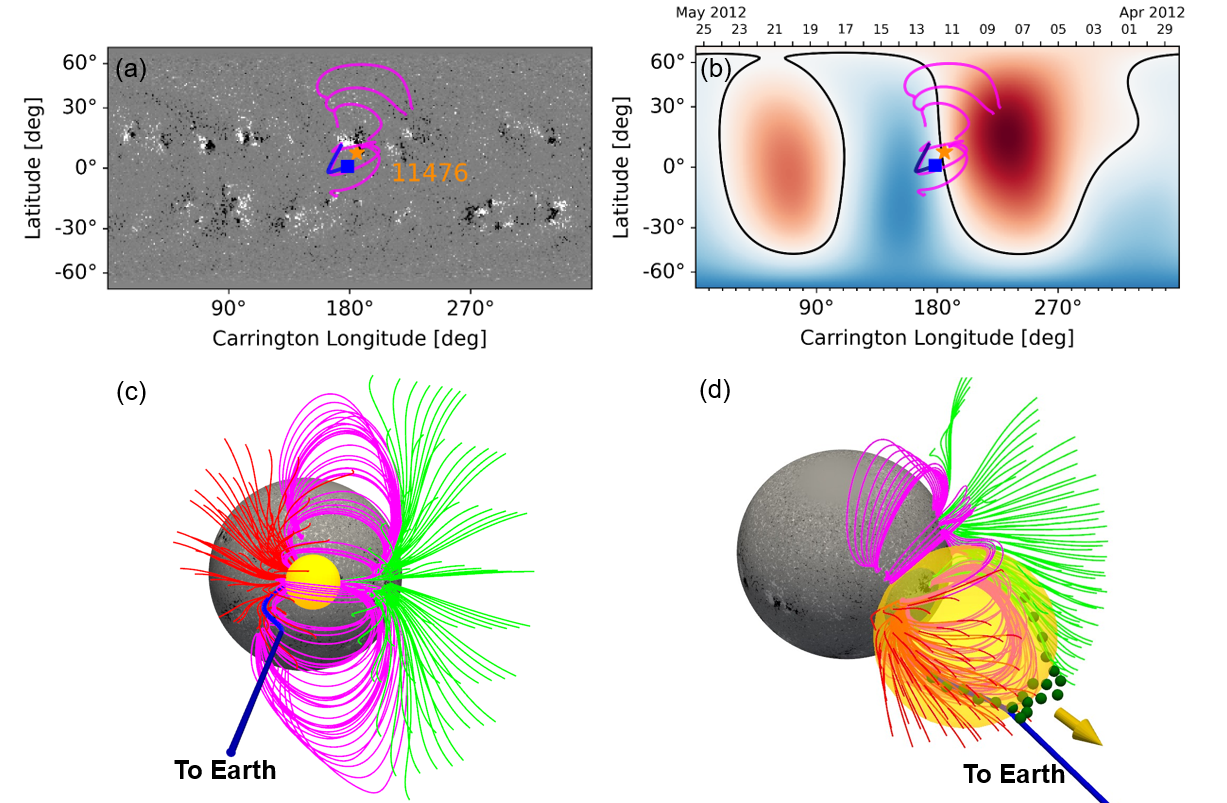}
\caption{Large-scale magnetic configuration around the AR11476 with selected field lines traced through the PFSS model. Closed field lines are plotted in magenta,
positive open field lines in red, and negative open field lines in green.
The footpoint of the Parker spiral connecting to the Earth at 2.5 R$_{\sun}$ is shown with the blue square, and the corresponding field line further traced down to the solar surface is also plotted in blue.
In panels (c) and (d), the yellow ellipsoids show the evolution of shock fronts at the time of shock formation and the
GLE release time when the shock apex reaches 2.4 R$_{\sun}$, respectively. The direction of the shock apex in panel (d) is indicated by an arrow in gold color. The small solid circles in panel (d) represent the highest-energy SEPs which are mainly accelerated at the top of closed loops and released when the shock moves through the streamer cusp region.
}
\label{fig:Bfield2012}
\end{figure}

For the GLE 71 event, Figure \ref{fig:Bfield2012} displays the large-scale magnetic configuration around the AR11476 with selected field lines traced through the PFSS model.
Closed field lines are plotted in magenta, positive open field lines in red, and negative open field lines in green.
In Figure~\ref{fig:Bfield2012}(a), the closed field lines resembling the streamer (belt) are overplotted on the HMI synoptic map for Carrington Rotation (CR) 2123, which is taken as the input for the PFSS modeling.
The location of the flare is indicated with a star.
This indicates that the flare occurred below the streamer and the CME-driven shock can interact closely with the streamer as it erupts and moves outward \citep{2016ApJ...833...45R}.
The footpoint of the Parker spiral connecting to the Earth at 2.5 R$_{\sun}$ is shown with the blue square, based on the average solar wind speed of $\sim$400 km s$^{-1}$ as measured by SOHO. The corresponding field line is further traced down to the solar surface, as plotted in blue, with its footpoint located near the eastern edge of the AR11476.
Figure \ref{fig:Bfield2012}(b) shows the synoptic source surface map at 2.5 R$_{\sun}$. The neutral line lying between hemispheres of outward-pointing (blue) and inward-pointing (red) magnetic fields is indicated with the black curve.
The magnetic footpoint of the Earth at the source surface (the blue square) is located near the neutral line.

Figures \ref{fig:Bfield2012}(c) and (d) show the evolution of the CME-shock, as represented by the yellow ellipsoid, combined with the surrounding magnetic field lines traced from the PFSS model, at the initial phase and at GLE particle release time, respectively.
As shown in Figure~\ref{fig:Bfield2012}(c), the CME emerged below the streamer belt.
The shock formation height is determined as $\sim$1.4 R$_{\sun}$ by \citet{2013ApJ...765L..30G} from the onset of the associated type II radio burst at 01:32 UT.
When the CME-shock is located below the streamer and enclosed by closed field lines, the shock geometry is mostly quasi-perpendicular \citep{2011ASTRA...7..387P,2016ApJ...821...32K,2016ApJ...833...45R}, therefore likely having a high particle acceleration rate \citep{2005ApJ...624..765G}. As simulated by \citet{2017ApJ...851...38K}, the highest-energy particles are preferentially accelerated at the top of closed field lines, where the upstream field lines intersect the shock at two points and a collapsing trap forms \citep{1993JGR....98...33D,2006A&A...455..685S}, producing ``hot-spots'' along the shock front \citep{2010ApJ...725..128G,2010ApJ...723..393K}.
This indicates that, in the GLE 71 event, the Earth can be well connected to the high-energy SEP source region in the early stage of the eruption, leading to rapid increase of SEP fluxes in all energy channels.
As shown in Figure~\ref{fig:Bfield2012}(d), at the GLE release time at 01:41 UT, the outermost shock front reached 2.4 R$_{\sun}$, as extrapolated from the 3D shock reconstruction.
This is consistent with the deduced CME-shock height at the GLE release time (2.32 R$_{\sun}$ at 01:40:26 UT) in \citet{2013ApJ...765L..30G}.
The high-energy SEPs accelerated at the top of closed field lines, as represented by small solid circles, can be released when the shock apex propagated through the streamer cusp (around 2.5 R$_{\sun}$) and encountered open field region.

\begin{figure}
\centering
\includegraphics[width=0.95\linewidth]{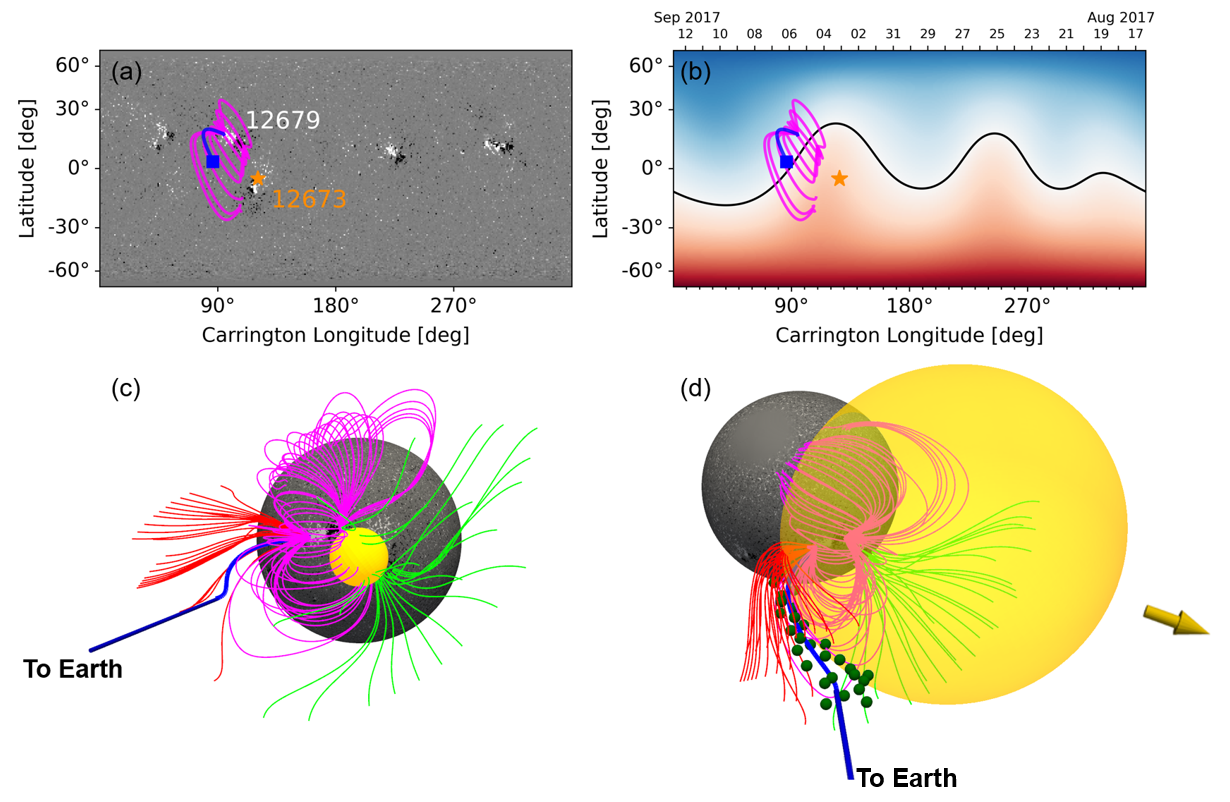}
\caption{Same as Figure \ref{fig:Bfield2012}, for the 2017 September 10 event.
In panel (d), the shock apex reaches 3.7 R$_{\sun}$ at the GLE particle release time, with the direction indicated with an arrow in gold color. The small solid circles illustrate the highest-energy SEPs released to the field line connecting to Earth.
}
\label{fig:Bfield2017}
\end{figure}

For the GLE 72 event, Figure \ref{fig:Bfield2017} shows the large-scale magnetic configuration around the AR12673.
The HMI synoptic magnetogram in CR2194 is used for PFSS modeling.
As shown in Figure~\ref{fig:Bfield2017}(a), the AR12673 where the flare and CME erupted is situated near the edge of the PFSS closed field, i.e., the streamer belt \citep{2018SpWea..16..557L}.
Thus, in this event, the CME-shock interacts with the streamer belt at the flanks during its expansion and propagation.
According to the in-situ solar wind speed of 450 km s$^{-1}$, the magnetic field line of the Earth is traced back to the source surface, with the magnetic footpoint (blue square) located near the neutral line, as shown in Figure~\ref{fig:Bfield2017}(b).
When we further trace the magnetic field line to the solar surface, the footpoint is connected to AR12679, located to the northeast of the eruption region AR12673 and below the closed field of the streamer belt.
Figures~\ref{fig:Bfield2017}(c) and (d) show the evolution of the CME-shock and the surrounding magnetic field lines traced from the PFSS model, at the initial phase and at GLE release time, respectively.
From the starting time of the associated type II radio burst at 15:53 UT, \citet{2018ApJ...863L..39G} deduced that the shock formed at a height of $\sim$1.4 R$_{\sun}$.
At the initial stage, the Earth is poorly connected with the CME shock and the SEP source region.
As the shock expands laterally, the shock flanks become magnetically connected to the Earth observer.
By linear interpolation from the 3D shock reconstruction, the shock apex moved to 3.7 R$_{\sun}$ at the GLE release time at 16:03 UT.
As shown in Figure~\ref{fig:Bfield2017}(d), the eastern shock flank was propagating through the streamer. The shock is nearly perpendicular when the shock flank interacts with the farther side of the streamer arcades.
The simulation in \citet{2019ApJ...883...49K} showed that the highest-energy particles are predominantly accelerated in the shock-streamer region due to quasi-perpendicular geometry and closed field lines.
Because the shock has to move some distance before interacting with the streamer at the flank and forming the configuration favorable for efficient particle acceleration, the height of the CME-shock is much higher at the GLE release time, compared to the case when the shock originates below the streamer.

By using 3D shock reconstruction and modeling, \citet{2020ApJ...893...76K} showed that at 16:00 UT the shock geometry is mainly quasi-perpendicular at the eastern flanks, while quasi-parallel at the shock apex. They also pointed out that at later time the shock flanks passed through the streamer regions where the shock strength can be enhanced due to a lower characteristic speed.
\citet{2018ApJ...863L..39G} showed that around the GLE release time, the eastern flank of the shock (as observed in EUV running difference images) was crossing the Sun-Earth field line, and also the continued metric type II burst indicates the source of energetic electrons is located at shock flanks (see also the LOFAR radio imaging observations by \citet{2019NatAs...3..452M}).
During the shock-streamer interaction process, particle acceleration gets more efficient as the shock angle is larger, and the magnetic connection between the high-energy SEP source to the Earth gets better, which may explain the hardening in proton spectrum shortly after the GLE onset time.

\section{Conclusions and Discussion}
In this paper, we analyze the two GLE events during solar cycle 24.
Important episodes are summarized in Table \ref{tab:event}.
We focus on the effects of large-scale coronal magnetic field configuration near the active regions on the acceleration and release of SEPs close to the Sun.

The GLE 71 event on 2012 May 17 was associated with an M5.1 class flare originating from the AR11476 at N13W87, and the GLE 72 event on 2017 September 10 was associated with an X8.2 class flare from the AR12673 at S09W91.
Although the solar eruptions both occurred near the western limb, the temporal evolution of SEP intensity profiles and energy spectra are different at the early stages of the two events.
In GLE 71, the proton intensities at above 100 MeV reached the peak and plateau phase promptly and then decay gradually, showing the typical behavior of a well-connected SEP event. Correspondingly, the energy spectra at high-energies quickly steepened into a power-law distribution with a spectral index of $\sim$3 within the first 40 minutes and remained almost unchanged.
In GLE 72, the proton intensities increased more slowly and took more than 2 hours before reaching the peak and plateau phase. The energy spectra at high-energies evolve more gradually into a power-law within the first 2 hours.
In addition, within the first $\sim$10 minutes after the GLE onset, the spectral index at $>$100 MeV is approximately a constant in GLE 71, but it decreases (spectral hardening) in GLE 72.

By combining EUV and white-light images from different viewpoints and PFSS extrapolation of the coronal magnetic field, we find that the CME in GLE 71 originated below the streamer belt, while in GLE 72 outside of the streamer belt.
The magnetic footpoint of the Parker spiral connecting to the Earth is located near the AR11476 in GLE 71, but to the northeast of the AR12673 in GLE 72.
We perform 3D reconstruction of the shock front using an ellipsoid model by fitting the coronagraph images from two different viewpoints. Then, we obtain the 3D shock front at the GLE particle release time by linear interpolation/extrapolation.

In GLE 71, since the CME emerged below the streamer belt, the shock front is surrounded by closed field lines and mostly quasi-perpendicular, leading to efficient particle acceleration.
Particularly, as simulated by \citet{2017ApJ...851...38K}, the highest-energy particles are predominantly accelerated at the shock nose, where a collapsing trap can form and produce ``hot-spots'' of energetic particles.
In the early stage, the Earth observer has connected well to the SEP source region, therefore rapid increase of SEP fluxes can be observed in all energy channels of GOES instruments.
At the GLE particle release time, the outermost shock front reached 2.4  R$_{\sun}$ and was crossing the streamer cusp region, where the Earth observer is magnetically connected.

In GLE 72, the CME-shock propagated through the streamer belt due to subsequent lateral expansion, thus particle acceleration is most efficient at the shock flanks, as simulated in \citet{2019ApJ...883...49K}.
Early in this event, the Earth is poorly connected to the SEP source region, which can explain the slow increase of SEP fluxes. A good connection became built later at the eastern shock flank, as the shock expanded farther and traveled to a larger altitude.
Similarly, at the GLE release time, the shock flank was sweeping across the streamer cusp region, where the Earth is magnetically connected. However, the shock nose has reached 3.7 R$_{\sun}$, a height much greater compared to that in GLE 71.
The spectral hardening at the first 10 minutes after the GLE onset may be due to increased acceleration efficiency during the shock-streamer interaction process.

Our analysis suggests that shock-streamer interactions may have affected the acceleration and release of the highest-energy particles in the early stages of the two GLE events.
Note that previous studies emphasized the importance of a supercritical shock for efficient particle acceleration \citep{2016ApJ...833...45R,2020ApJ...893...76K}. However, \citet{2022ApJ...926..227F} showed that a subcritical shock can also accelerate particles to $>$100 MeV as a result of prolonged shock-streamer interaction.
As shown in previous statistical analysis of GLE events \citep[e.g.,][]{2009ApJ...706..844R,2009ApJ...693..812R,2012SSRv..171...23G}, the height of the CME/shock leading edge at the GLE release time depends on the source longitude (or the longitude relative to the footpoint of the field line connecting to the observer), much lower for well-connected events compared to poorly-connected events.
However, it can also be found that the CME/shock heights can differ largely, even for events in similar source longitude.
As revealed in our study, it may be necessary to take the effects of large-scale magnetic configuration near the active regions into account. Magnetic structures such as the streamers can affect the particle acceleration efficiency at the shock and therefore the source location of the highest-energy particles.
Note that pseudo-streamers with closed fields usually below 1.5 R$_{\sun}$ may also work when the shock can form low in the corona and interact closely with the pseudo-streamers.
To provide more evidence from observations, more GLE and large SEP events will be investigated in future work.
In addition, it is worthwhile to explore whether or not the shock-streamer interaction plays a role in the formation of GLEs in comparison with non-GLEs.

\begin{acknowledgments}
X.K. thanks Dr. Ryun Young Kwon for helpful discussions on the ellipsoid shock model.
This work was supported by the National Natural Science Foundation of China under grants 42074203 and 11873036, Yunnan Key Laboratory of Solar Physics and Space Science under grant YNSPCC202218, the Young Elite Scientists Sponsorship Program by China Association for Science and Technology, and the Young Scholars Program of Shandong University, Weihai.
F.G. is supported by NSF grant AST-2109154 and DOE grant DE-SC0018240.
\end{acknowledgments}

%

\vspace{5mm}
\facilities{GOES, SDO, SOHO, STEREO, Neutron Monitors (FSMT, OULU)}


\software{Sunpy \citep{2020ApJ...890...68S}, pfsspy \citep{2020JOSS....5.2732S},
          PyThea \citep{2022FrASS...9.4137K}
          }








\bibliography{export-bibtex}{}
\bibliographystyle{aasjournal}


\end{document}